\documentstyle[aps,prl,epsf,floats]{revtex} 
\draft

\begin{document}
\twocolumn[\hsize\textwidth\columnwidth\hsize\csname
@twocolumnfalse\endcsname 

%\begin{flushright}
%UCLA/01/TEP/04 
%BNL-HET-01/** 
%\end{flushright}

\preprint{UCLA/01/TEP/04;  BNL-HET-01/21
} 

\title{Q-ball candidates for  self-interacting dark matter
} 

\author{Alexander Kusenko$^{1,2}$ and Paul J. Steinhardt$^3$
}
\address{$^1$Department of Physics and Astronomy, UCLA, Los Angeles, CA
90095-1547 \\ $^2$RIKEN BNL Research Center, Brookhaven National
Laboratory, Upton, NY 11973 \\
$^3$Department of Physics, Princeton University, Princeton, NJ 08540
}

\date{April, 2001}

\maketitle
             
\begin{abstract}

We show that non-topological solitons, known
as Q-balls, are promising candidates for self-interacting
dark matter.  They can satisfy the cross-section requirements for a
broad range of masses.  Unlike previously considered examples,
Q-balls  can stick together after collision, reducing the effective
self-interaction rate to a negligible value after
a few collisions per particle.  This feature modifies predictions
for halo formation.
We also discuss the possibility
that Q-balls  have large interaction cross-sections
with ordinary matter.

\end{abstract}

\pacs{PACS numbers: 95.35.+d, 98.80.-k, 98.80.Cq  \hspace{1.0cm}
BNL-HET-01/21; UCLA/01/TEP/04} 

\vskip2.0pc]

\renewcommand{\thefootnote}{\arabic{footnote}}
\setcounter{footnote}{0}

The standard cold dark matter model based on non-relativistic,
collisionless particles
successfully predicts the formation of structure on large
scales exceeding a megaparsec, but appears to make problematic predictions
about structure on galactic and sub-galactic scales.
The dark matter density profile  in the cores of  galaxies, the
number of satellites, the thickenings of disks, the density of low
mass objects, gravitational lens statistics, and the
asphericity of cluster cores found in numerical simulations
appear to be at variance with observations~\cite{sidm1,sidm3}. 
The difficulties  suggest either that dark matter is not cold\cite{warm}
or that
dark matter is not collisionless~\cite{sidm1}.    
In either case, the conventionally
preferred candidates for dark matter,
weakly interacting massive particles (WIMPS) or axions, would be ruled out.

In this paper, we propose non-topological solitions known as Q-balls
as candidates for collisional dark matter.  
Q-balls occur in a wide range of particle physics models, can be
produced copiously in the early universe, and can be stable.
As candidates for self-interacting dark matter, they possess 
particularly
advantageous and interesting features, as shown in this paper.  
First, Q-balls can satisfy the requisite cross-section conditions 
for a much wider range of masses than ordinary point-like particles.
Second,  depending on the detailed interactions of the fields from 
which they are generated, Q-balls can scatter  
inelastically, leading to modifications of halo evolution compared
to elastically scattering collisional  dark matter.  
A particularly interesting limit is where they collide and  stick.
In this case, the population evolves and the cross-section and mass 
relations change as scattering proceeds. The effect is
that scattering becomes insignificant
as time proceeds.  The initial scatterings
 smooth out halo cores, but 
heat conduction ceases afterwards and gravothermal collapse is avoided.
A third feature of Q-balls is that they can
have significant interactions with ordinary matter (although this
is not required).  A large range of 
parameter space is ruled out by current experimental bounds, but
significant unconstrained range remains, suggesting new directions
for dark matter searches.

\noindent
{\it Basic Properties of Q-balls}
Q-balls are the ground state configurations for fixed charge $Q$
in  theories with interacting scalar fields $\phi$ that carry
some conserved global U(1) charge~\cite{q,coleman,nts_review}.
If the  field configuration is written in the form
\begin{equation}
\phi(x,t) = e^{i \omega t} \bar{\phi}(x),
\label{q}
\end{equation}
its  charge is
\begin{equation}
Q= \frac{1}{2i} \int \phi^* \stackrel{\leftrightarrow}{\partial}_t  
\phi \, d^3x = \omega \int \bar{\phi}^2 d^3x . 
\label{Qt}
\end{equation}
The form of $\bar{\phi}(x)$ is determined minimizing the energy
\begin{equation}
E=\int d^3x \ \left [ \frac{1}{2} |\dot{\phi}|^2+
\frac{1}{2} |\nabla \phi|^2 
+U(\phi) \right],
\label{e}
\end{equation} 
where the potential $U(\phi)$ has a minimum at $\phi=0$ 
and is invariant under the global U(1) transformation
$\phi \rightarrow e^{i \theta} \phi$.
In the thin-wall limit 
$\bar{\phi}(x) =
\phi_0$ is nearly constant in the interior ($r <R$) and drops 
rapidly to zero for $r>R$.  
Using  (\ref{Qt}), one can write the
energy~(\ref{e}) as 
\begin{equation}
E \approx \frac{Q^2}{2 V \phi_0^2} + VU(\phi_0),  
\label{ea}
\end{equation} 
where $V$ is the Q-ball volume.  The minimum of energy in eq.~(\ref{ea})
with respect to $V$ is $E=\mu Q$, where $\mu = \sqrt{ 2 U(\phi_0) /\phi_0^2
}$. In the thin-wall limit, the minimum of $E$ 
with respect to the value $\phi_0$ corresponds to
\begin{eqnarray}
\mu \rightarrow  \mu_0 & = & min\left(\sqrt{ \frac{2 U(\phi)}{\phi^2}}\right).
\label{condmin}
\end{eqnarray}

Depending on the potential, $\mu_0$ in eq.~(\ref{condmin}) can be finite
(Type~I) or infinite (Type~II).  
The mass of a Type~I Q-ball  is
$M(Q) = \mu Q$
For large Q-balls ($Q\rightarrow \infty$), $\mu\rightarrow \mu_0$
and $\phi \rightarrow \phi_0$ 
(``thin-wall'' limit)~\cite{q,coleman,nts_review}.  For smaller values of
$Q$, $\mu$ can be computed in a ``thick-wall''
approximation~\cite{ak_qb}.  For $Q<10$ radiative corrections become
important~\cite{graham}.  
In any case, $\mu$ is less than the
mass of the $\phi$ particle as a consequence of condition (\ref{condmin}).
The radius of the Type~I Q-ball is 
\begin{equation}
R_{_Q} \approx \left ( \frac{3}{4 \pi}\right )^{1/3} \frac{ Q^{1/3} }{(\mu
\phi_0^2)^{1/3}} .
\end{equation}

Type~II Q-balls
occur if the scalar potential grows slower than the second power
of $\phi$. Then the Q-ball never
reaches the thin-wall regime, even if $Q$ is large.  The value of $\phi$
inside the soliton extends to a value as large as the gradient terms allow,
and the 
mass of a Q-ball is proportional to $Q^{p}$, $p<1$.  In particular, if the
scalar potential has a flat plateau $U(\phi) \sim m_{\rm flat}^4 $ at large
$\phi$, then the mass of a Q-ball is~\cite{dks}
$M(Q) \sim m_{\rm flat} Q^{3/4}$
and the   size is  $R_{_Q} \sim m_{\rm flat}^{-1}
Q^{1/4}$.

\noindent
{\it Q-balls as self-interacting dark matter:}
As in generic examples of
self-interacting dark matter, Q-ball scattering in regions of high 
density facilitates heat exchange  in dark matter cores that smooths
out their distribution and, also, enhances  the stripping of dark 
matter from satellites that accelerates their tidal destruction. 
Both effects serve to resolve the problems of cold, collisionless
dark matter.
For  these purposes, the basic requirement
is that the ratio of self-interaction cross section $\sigma_{_{DD}}$ to
particle mass $M$ must be in the range\cite{sidm1}
\begin{eqnarray}
S=\frac{\sigma_{_{DD}}}{M} & = & 8\times 10^{-25}- 1\times10^{-23}\, 
{\rm cm}^2\,{\rm GeV}^{-1} \nonumber \\
& = & 0.5-6\, {\rm cm}^2\,{\rm g}^{-1},  
\label{sigma}
\end{eqnarray}
For point particles whose dominant scattering is $s$-wave, 
Hui has shown  that unitarity implies
a cross-section  bounded above by
$\sigma_{DD} \sim 1/(M v_{rel})^2$,
where $v_{rel} \approx 300$~km/s is the typical velocity of the 
dark matter particles. In this case, the maximal mass
for a point particle is $M \approx 10$~GeV~\cite{Hui}.  
Q-balls are extended objects which can evade this bound.  
Higher partial waves contribute to their scattering such that 
their cross-section is essentially geometric (except in the 
limit of very small coupling), $\sigma_{DD} = \pi R_Q^2$.
Then, $S$ for Type~I Q-balls  is
\begin{equation}
\frac{\sigma_{_{DD}}}{M} \approx \left (  \frac{9 \pi}{16} \right )^{1/3} 
\left ( \frac{M(Q) }{\phi_0} \right )^{4/3} \ 
\frac{Q^{4/3}}{M(Q)^3},
\label{sigma_Q}
\end{equation}
and for Type~II Q-balls is
\begin{equation}
\frac{\sigma_{_{DD}}}{M} \sim m_{\rm flat}^{-3} Q^{-1/4} \sim
\frac{Q^2}{M(Q)^3} .
\label{sigma_Q_flat}
\end{equation}
Note that both expressions for $S$ can greatly exceed the unitarity
bound for large $Q > 10^5$.  Hence, it is possible to have $Q$-ball
candidates that satisfy the requirements on $S$ for a range of masses
much greater than 10~GeV (the unitarity limit).

If  Q-balls are of Type I and no restriction is placed on the relative
magnitude of $\phi_0$ and $\mu$, the mass of the $\phi $ particle can range
from below a keV to well beyond the electroweak scale.  If the mass of
$\phi > M_Z$, such a scalar field could make extremely heavy, strongly
interacting Q-balls ({\em cf.}  Ref.~\cite{kkst}).

Naturalness arguments, while not rigorous,
suggest potentials in which
$\phi_0 \sim \mu$.  In this case, for Type I Q-balls ($M(Q)\sim \mu Q$), a
satisfactory choice of parameters is in the range around $\mu\sim \phi_0
\sim 20$MeV, $Q\sim 10-10^3$.  
For Type II Q-balls (with 
$M(Q)\sim m_{\rm flat} Q^{3/4}$), the analogous relations
are $m_{\rm flat} \sim
20$~MeV and  $Q\sim 10^4-10^5$.  
%CHANGE
Note that,
if the global U(1) symmetry of the Q-balls is associated with
baryon number, as in most examples considered previously~\cite{ks},
 empirical constraints specific to 
 baryonic processes do not permit   the requisite 
large cross-sections.  However, there is no problem with 
more general U(1) symmetries.

\noindent
{\it Q-ball production in the early universe:}
Several mechanisms could lead to a formation of Q-balls in the early
universe. First, they can be produced in the course of a phase
transition~\cite{s_gen}.  Second, solitosynthesis, a process of  gradual
charge accretion similar to nucleosynthesis, can take
place~\cite{gk,ak_pt,dew}.  Finally, Q-balls can emerge from 
fragmentation of a scalar condensate~\cite{ks} formed at the end of
inflation.

Solitosynthesis occurs through an accretion of charge.  It requires
some universal asymmetry $\eta_{_Q}$ of the global charge $Q$, similar to
baryon asymmetry of the universe.  
When the temperature drops below some critical value $T_c \sim \left
( \mu/|\ln \eta_{_Q}|\right )$~\cite{ak_pt}, a Q-ball minimizes
both the energy and the free energy of the system, and a rapid coalescence
of global charge into Q-balls occurs~\cite{gk,ak_pt}.  The number of
Q-balls and their mass density in the universe depends on the value of
$Q$-asymmetry, $\eta_{_Q}$, and is largely unconstrained.

Fragmentation of a coherent scalar condensate can
lead to a copious production of Q-balls~\cite{ks}.  At the end of
inflation, scalar fields develop large expectation values along those
directions in the potential that have small masses or flat
plateaus~\cite{ad}.  The subsequent rolling of the condensate can encounter
an instability, as a result of which the scalar condensate can break up
into Q-balls~\cite{ks}.  This process has been studied both
analytically~\cite{ks,em} and numerically~\cite{kasuya,num} and was shown
to produce a sharply peaked distribution of sizes of Q-balls.  There is
also some evidence that Q-balls and anti-Q-balls can form from the same
condensate while the overall charge asymmetry $\eta_{_Q}$~\cite{num} is
small or zero.  The number density of Q-balls formed in this way depends on
the shape of the potential at large $\phi$ and the horizon size at the time of
formation.  The only strict constraint is that the separation between
Q-balls should be of the order of their size at the time of formation.  For
us this translates into a red shift at which Q-balls are formed.

\noindent
{\it Q-ball scattering and sticking:} Whereas previous studies of
collisional dark matter assumed that they scatter elastically, Q-balls can
either merge or split after a collision depending on whether energy can be
dissipated~\cite{collisions}.  The merger of Q-balls requires the kinetic
energy to dissipate quickly on the time scale of the collision. In the
absence of additional interactions, emission of $\phi$ particles is the
only channel for such dissipation.  There are many modes of oscillations of
an excited Q-ball: volume, surface, etc. The basic mode for a Type I Q-ball
deforms the entire Q-ball and has a frequency $\Omega \sim \phi_0^{2/3}
\mu^{1/3} Q^{-1/3}$, which decreases with $Q$.  Production of $\phi$
particles in these oscillations is very efficient if $\Omega > m_\phi$.
When $\Omega$ is close to $m_\phi$, production of $\phi$ particles is
enhanced by parametric resonance.  For $\Omega > m_\phi$, there are several
other resonant bands.  If, however, $\Omega < m_{\phi}$, particles are
produced very inefficiently, and it is unlikely that the Q-ball will
dissipate any energy at all on the time scale of a typical collision.
Therefore, if $\Omega > m_\phi$, Q-balls can merge, but if $\Omega <
m_\phi$, they are more likely to fragment.  Additional interactions of
$\phi$ with other light states can enhance the dissipation and increase the
probability of merging.  For Type-II Q-balls, there is a strong energetic
bias toward merging as opposed to fragmentation.  The subject of Q-ball
collisions, clearly, deserves further studies.  

Merger or sticking together of dark matter can lead to novel dynamics
of the halo compared to the standard case of elastic self-interactions.
As perturbations begin to grow, the density of Q-balls is too low 
for there to be significant scattering.  As the halo density
profile becomes
steeper and denser, Q-ball collision and merger takes place.   Mergers
replace two particles with charge $Q$ with a single particle of charge
$2Q$.  According to  Eqs.~(\ref{sigma_Q}) and~(\ref{sigma_Q_flat}), 
the ratio $S \equiv \sigma_{DD}/M$ decreases as Q increases
%CHANGE
and the mean velocity $v_{rel}$  decreases by two.
Both effects decrease the interaction rate until, ultimately, 
self-interaction ceases.
%ENDCHANGE
Thereby, a stable population of Q-balls is reached with lower
central density than occurs if there are no collisions.
A modest amount of kinetic energy is lost, but the fraction 
of particles that scatter (halo particles scattering off particles
in the central core) is a tiny fraction of the total halo.  
Because the collisions self-adjust the population from interacting
to non-interacting, heat conduction ceases and gravothermal 
collapse is avoided.    Hence, it is interesting to compare predictions
for small halos formed early in the universe when the density is high.
For collisionless dark matter, there are many such halos and they are
extremely cuspy and dense. For elastically scattering dark matter, collisions
smooth the core distribution but, then, gravothermal collapse
causes the central density to rise again.  In the case of merging 
Q-balls, the core density is reduced  and it remains that way.

In the case of splitting, binding energy can be converted into 
kinetic energy. Since the binding energy can exceed the gravitational
binding to the halo, splitting  can lead to conversion of two
similar-size Q-balls into one large Q-ball that remains gravitationally
bound to the halo and one small fragment that escapes.
It is possible to imagine that both merger and splitting play a role.
Suppose $S \equiv \sigma_{DD}/M$ is initially large. In a small young
halo with a dense core, collision and merger transforms the population
into large Q-balls with small $S$.  Large Q-balls are more likely to 
split and have energy escape.  The halo structure will be influenced
by both effects.

\noindent
{\it 
What kinds of interaction are possible between Q-balls and baryons?}
If the field $\phi$ has only gravitational interactions with matter, 
Q-balls cannot be detected directly. 
 Condition~(\ref{sigma})  for  $\phi_0 \sim \mu$ (the naturalness
 condition)
suggests that
$m_\phi$ must be close to
$0.02$GeV.  
But, if $\phi$ has a mass below 50 MeV, it
cannot have strong
interactions with matter because of a combination of collider bounds, in
particular because a neutral pion could decay into $\phi \bar{\phi}$ pairs.
Likewise, a light $\phi$ field cannot have weak interactions because it
would violate the precision measurements at LEP of the $Z$ width.  

Q-balls can interact strongly with ordinary matter if $m_\phi > 
50$~MeV.  Alternatively, there is the 
the possibility of a  significant cross-section with ordinary matter
if the  interactions are mediated by some  new physics.
In a number of Grand
Unified and string-inspired models, light fields are accompanied by
additional interactions.  For example, an interesting possibility is that
an additional gauge U(1)', unrelated to either the global U$_{_Q}$(1) or
the Standard Model gauge group, is spontaneously broken at some high
scale~\cite{Zprime}.  We will use this model to illustrate possible
interactions that the
field $\phi$ can have with matter, which can ultimately
make Q-balls detectable.
Let us suppose that $\phi$ interactions with matter are mediated by some
vector boson with mass $M_{Z'}$.  Then the cross section for Q-ball
interactions with a nucleon is, roughly,   
\begin{equation}
\sigma_{Qp} \sim F \frac{g^2 Q^2}{M^2_{Z'}},  
\end{equation}
where $g$ is some coupling constant and $F$ is a form factor.  The
$Q^2$-dependence occurs because of the coherent scattering of a nucleus off
the $\phi$ quanta in the condensate, and the form factor $F$ accounts for a
fraction of Q-matter that scatters coherently.  If the size of the Q-ball
is smaller than that of a nucleus, $F$ is of the order of 1.  If the Q-ball
is much larger than the nucleus, $F\sim (R_n/R_{_Q})^3$, where $R_n$ is the
size of  the nucleus.

\begin{figure}[t]
\centering
\hspace*{-5.5mm}
\leavevmode\epsfysize=6.5cm \epsfbox{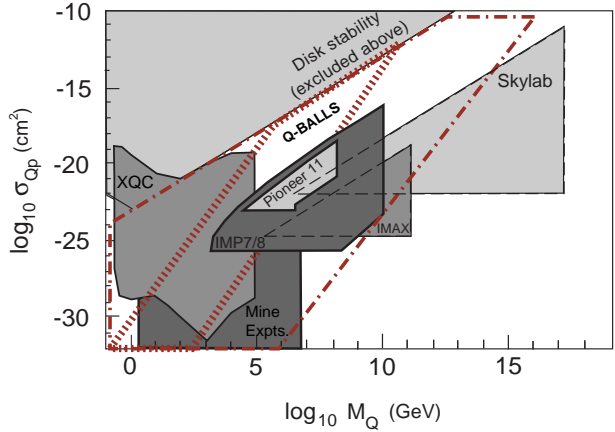}\\[3mm]
\caption[fig1]{\label{fig1}
%CHANGE
Empirical constraints (shades of grey)  on Q-ball/proton
cross section and mass assuming $\phi$ interacts through an
intermediate boson $Z'$ with
$M_{Z'} \approx 1$~TeV and $g \approx 0.1$. Experiments are described in
Ref.~\cite{last}.   The most likely range for Q-balls
is enclosed by hatched boundary
(either Type I with $\phi_0/\mu$ within a few orders of 
magnitude of unity or Type II).
White regions are currently unconstrained experimentally.
If no restriction is placed on
$\phi_0/\mu_0$, the predicted range for Q-balls 
expands to the dot-dashed boundary, including even more untested
territory.
}                
\end{figure}

The resulting dark matter (Q-ball)/proton cross-ections, $\sigma_{Qp}$,
are large enough to be detected for a broad
range of parameters.  Fig.~1 summarizes the current limits on 
$\sigma_{Qp}$ and $M$ based on existing searches~\cite{sidm3}.
Superposed are the predictions for Q-balls.  A large range of parameters
is already ruled out, but there remain unexplored regimes.
One  consists of Q-balls with
large cross-sections and masses larger than 
a TeV. Since the local dark matter density is 0.4~GeV/cm$^3$ and 
the mean velocity is 300~km/s, the flux is of order
$10^{5} {\rm cm}^{-2} {\rm s}^{-1}$.
Another possibility is relatively light Q-balls with masses about 1~GeV
and a weak cross section, below $10^{-31}{\rm cm}^{2}$.  The flux of these
particles would be high,  of order  $10^{10} {\rm cm}^{-2} {\rm s}^{-1}$.
Different strategies would have to be adopted to search in the two regimes,
but both are feasible, as will be discussed in a future paper.

To summarize, a new scalar field can, in the form of Q-balls, be the cold
dark matter consistent with all present observations.  The
self-interactions of Q-balls are characterized by a large cross section due
to their extended geometry, a property that can naturally explain
the flattened density profiles of dark matter halos.
If Q-balls scatter inelastically and merge, scattering may cease
after the profile is flattened.   It is conceivable that 
Q-balls have significant interactions with ordinary matter,
either  strong interactions
or  interactions mediated by a heavy $Z'$ boson.  
In this case, the Q-balls 
 can be detected in near-future experiments.

We thank D. Spergel and J.P. Ostriker for many 
useful remarks,
G. Gelmini and S. Nussinov for discussions of Q-ball interactions
with matter, and P. McGuire for aid in determining existing constraints
in the Figure.
This work was supported by in part by Department of Energy
grants DE-FG03-91ER40662 (AK) and DE-FG02-91ER40671 (PJS).


\begin{thebibliography}{99}

\bibitem{sidm1}
D.~N.~Spergel and P.~J.~Steinhardt,
%``Observational evidence for self-interacting cold dark matter,''
Phys.\ Rev.\ Lett.\ {\bf 84}, 3760 (2000)
[astro-ph/9909386].
%%CITATION = ASTRO-PH 9909386;%%

\bibitem{sidm3} For an overview, see, for example,
B.~D.~Wandelt, R.~Dave, G.~R.~Farrar, P.~C.~McGuire,
D.~N.~Spergel and P.~J.~Steinhardt,
%``Self-interacting dark matter,''
astro-ph/0006344.
%%CITATION = ASTRO-PH 0006344;%%

\bibitem{warm}  J. Dalcanton and C. Hogan, astro-ph/0004381;
P. Bode, J.P. Ostriker, and N. Turok,
astro-ph/0010389.






\bibitem{q} G.~Rosen, J. Math. Phys. {\bf 9}, 996 (1968); 
{\em ibid.} {\bf 9}, 999 (1968);   
R.~Friedberg,  T.~D.~Lee, A.~Sirlin,  Phys. Rev. D {\bf 13}, 2739 (1976).
%%CITATION = PHRVA,D13,2739;%%

\bibitem{coleman} S.~Coleman, Nucl. Phys. B {\bf 262}, 263 (1985).
%%CITATION = NUPHA,B262,263;%%

\bibitem{nts_review}
T.~D.~Lee and Y.~Pang,
%``Nontopological solitons,''
Phys.\ Rept.\  {\bf 221}, 251 (1992).
%%CITATION = PRPLC,221,251;%%

\bibitem{ak_qb} 
A.~Kusenko, 
%``Small Q balls,''
Phys.\ Lett.\  B {\bf 404}, 285 (1997). 
%%CITATION = HEP-TH 9704073;%%

\bibitem{graham}
N. Graham, 
%``Quantum corrections to Q-balls,''
hep-th/0105009.
%%CITATION = HEP-TH 0105009;%%

\bibitem{dks} 
G.~Dvali, A.~Kusenko, M.~Shaposhnikov,
%``New physics in a nutshell, or Q-ball as a power plant,''
Phys.\ Lett.\  B {\bf 417}, 99 (1998).
%%CITATION = HEP-PH 9707423;%%

\bibitem{Hui} L. Hui, astro-ph/0102349.

\bibitem{ks} 
A.~Kusenko, M.~Shaposhnikov,
%``Supersymmetric Q-balls as dark matter,''
Phys.\ Lett.\  B {\bf 418}, 46 (1998).
%%CITATION = HEP-PH 9709492;%%

\bibitem{kkst} A.~Kusenko, V.~Kuzmin, M.~Shaposhnikov and P.~G.~Tinyakov,
%``Experimental signatures of supersymmetric dark-matter Q-balls,''
Phys.\ Rev.\ Lett.\ {\bf 80}, 3185 (1998)
[hep-ph/9712212].
%%CITATION = HEP-PH 9712212;%%

\bibitem{s_gen} J.~A.~Frieman, G.~B.~Gelmini, M.~Gleiser, E.~W.~Kolb, 
Phys. Rev. Lett. {\bf 60}, 2101 (1988);  
K.~Griest, E.~W.~Kolb, A.~Maassarotti, Phys. Rev. D {\bf 40}, 3529 (1989);  
J.~Ellis, J.~Hagelin, D.~V.~Nanopoulos, K.~Tamvakis, Phys. Lett. B {\bf
125}, 275 (1983).

\bibitem{gk} 
K.~Griest, E.~W.~Kolb, 
%``Solitosynthesis: Cosmological Evolution of Nontopological Solitons,''
Phys.\ Rev.\  D {\bf 40}, 3231 (1989); 
%%CITATION = PHRVA,D40,3231;%%
J.~A.~Frieman, A.~V.~Olinto, M.~Gleiser, C.~Alcock,
%``Cosmic Evolution Of Nontopological Solitons. 1,''
Phys.\ Rev.\  D {\bf 40}, 3241 (1989). 
%%CITATION = PHRVA,D40,3241;%%


\bibitem{ak_pt} 
A.~Kusenko,
%``Phase transitions precipitated by solitosynthesis,''
Phys.\ Lett.\  B {\bf 406}, 26 (1997).
%%CITATION = HEP-PH 9705361;%%

\bibitem{dew}
S.~Khlebnikov, I.~Tkachev, 
%``Quantum dew,''
Phys.\ Rev.\  {\bf D61}, 083517 (2000).
%%CITATION = HEP-PH 9902272;%%



\bibitem{ad} I.~Affleck, M.~Dine, Nucl. Phys. B {\bf  249}, 361 (1985); 
%%CITATION = NUPHA,B249,361;%%
M.~Dine, L.~Randall, S.~Thomas,  Phys. Rev. Lett. {\bf 75}, 398 (1995);
Nucl. Phys. B {\bf  458}, 291 (1996);  
%%CITATION = HEP-PH 9507453;%% %%CITATION = HEP-PH 9503303;%%
A.~Anisimov and M.~Dine, 
%``Some issues in flat direction baryogenesis,''
hep-ph/0008058.
%%CITATION = HEP-PH 0008058;%%

\bibitem{em} K.~Enqvist, J.~McDonald, 
%``Q-balls and baryogenesis in the MSSM,''
Phys.\ Lett.\  B {\bf 425}, 309 (1998); 
%%CITATION = HEP-PH 9711514;%%
%``B-ball baryogenesis and the baryon to dark matter ratio,''
Nucl.\ Phys.\  B {\bf 538}, 321 (1999);
%%CITATION = HEP-PH 9803380;%%
%``D-term inflation and B-ball baryogenesis,''
Phys.\ Rev.\ Lett.\  {\bf 81}, 3071 (1998);
%%CITATION = HEP-PH 9806213;%%
%``MSSM dark matter constraints and decaying B-balls,''
Phys.\ Lett.\  B {\bf 440}, 59 (1998);
%%CITATION = HEP-PH 9807269;%%
%K.~Enqvist and J.~McDonald,
%``Observable isocurvature fluctuations from the Affleck-Dine condensate,''
Phys.\ Rev.\ Lett.\  {\bf 83}, 2510 (1999)
%%CITATION = HEP-PH 9811412;%%
%``The dynamics of Affleck-Dine condensate collapse,''
%%CITATION = HEP-PH 9908316;%%
%K.~Enqvist, A.~Jokinen and J.~McDonald,
%``Flat direction condensate instabilities in the MSSM,''
Phys.\ Lett.\  {\bf B483}, 191 (2000). 
%%CITATION = HEP-PH 0004050;%%


\bibitem{kasuya} 
S.~Kasuya, M.~Kawasaki, 
%``Q-ball formation through Affleck-Dine mechanism,''
Phys.\ Rev.\  D {\bf 61}, 041301 (2000); 
%%CITATION = HEP-PH 9909509;%%
S.~Kasuya and M.~Kawasaki, 
%``Q-ball formation in the gravity-mediated SUSY breaking scenario,''
Phys.\ Rev.\  {\bf D62}, 023512 (2000).
%%CITATION = HEP-PH 0002285;%%

\bibitem{num}
K.~Enqvist, A.~Jokinen, T.~Multamaki and I.~Vilja,
%``Numerical simulations of fragmentation of the Affleck-Dine condensate,''
hep-ph/0011134.
%%CITATION = HEP-PH 0011134;%%


\bibitem{collisions}
M.~Axenides, S.~Komineas, L.~Perivolaropoulos, M.~Floratos,
%``Dynamics of nontopological solitons: Q balls,''
Phys.\ Rev.\  {\bf D61}, 085006 (2000);
%%CITATION = HEP-PH 9910388;%%
R.~Battye and P.~Sutcliffe,
%``Q-ball dynamics,''
hep-th/0003252;
%%CITATION = HEP-TH 0003252;%%
T.~Multamaki and I.~Vilja,
%``Q-ball collisions in the MSSM: Gravity-mediated supersymmetry breaking,''
Phys.\ Lett.\  {\bf B482}, 161 (2000);
%%CITATION = HEP-PH 0003270;%%
T.~Multamaki and I.~Vilja,
%``Q-ball collisions in the MSSM: Gauge-mediated supersymmetry breaking,''
Phys.\ Lett.\  {\bf B484}, 283 (2000).
%%CITATION = HEP-PH 0005162;%%


\bibitem{Zprime}
P.~Langacker and M.~Luo,
%``Constraints on additional Z bosons,''
Phys.\ Rev.\ D {\bf 45}, 278 (1992).
%%CITATION = PHRVA,D45,278;%%
J.~Erler and P.~Langacker,
%``Constraints on extended neutral gauge structures,''
Phys.\ Lett.\ B {\bf 456}, 68 (1999)
[hep-ph/9903476].
%%CITATION = HEP-PH 9903476;%%

\bibitem{last} P.C. McGuire and P.J. Steinhardt,  astro-ph/0105567.

\end{thebibliography}
\end{document}